\documentclass{article}

\usepackage{epsfig}

\begin{document}

\title{The importance of inelastic channels in eliminating continuum ambiguities in pion-nucleon partial wave analyses\footnote{\uppercase{T}his is an invited plenary talk held at the \uppercase{T}he \uppercase{P}hysics \uppercase{O}f \uppercase{E}xcited \uppercase{B}aryons (\uppercase{NSTAR} 2005) conference in \uppercase{T}allahassee, \uppercase{F}lorida, 12 - 15 \uppercase{O}ctober 2005, the full size conference presentation is located at \newline 
http://hadron.physics.fsu.edu/NSTAR2005/TALKS/Wednesday/Plenary/Svarc.ppt}}

\author{ A. \v{S}varc\footnote{\emph{E-mail address:} Alfred.Svarc@irb.hr}, S. Ceci and B. Zauner\\ \emph{Rudjer Bo\v{s}kovi\'{c} Institute}\\ \emph{Bijeni\v{c}ka c. 54}\\ \emph{10 000 Zagreb, Croatia}}

\maketitle

\section{Continuum ambiguities in partial wave \\analyses}
In spite of widespread interest in the methods of partial wave analyses (PWA), the concept of "continuum ambiguities", 
which is closely connected to the analysis of processes in the inelastic region, is nowadays rarely discussed. 
However, the oldest and very frequently referenced PWAs\cite{Cut79,KH84} do mention it, but they approach the problem 
in a particularly cautious way. In Cutkosky et.al.\cite{Cut79} the method of stabilizing the PWA solutions using 
hyperbolic dispersion relations is utilized, while in H\"{o}hler\cite{KH84}, at the very beginning of his monography, 
a whole sub-chapter is devoted to introducing the problem. In these paragraphs it is carefully argued that the 
elaborated method of imposing fixed-t analyticity together with isospin invariance is sufficient to produce the unique 
solution. Both analyses are unfortunately quite vague about the origin of the problem, and the very details of the 
technical aspect of utilized stabilizing procedures are given only in principle. In order to get the feeling how 
popular the problem of continuum ambiguities is, we have used the Google WEB search looking for places and times when 
it has been mentioned. Surprisingly, nothing much after mid 70-es has been found, and at that time the problem has 
been discussed mostly in mathematical literature\cite{Atk73,Bow75}. So, we believe that we are bound to refresh the 
basic knowledge on the issue. 
\section{What does it mean "continuum ambiguity"?}
The differential cross section itself is not sufficient to determine the scattering amplitude, because if $d\sigma / 
d\Omega = |F|^2$, then the new function \emph{\~{F}}~$= e^{i\Phi} F$ gives exactly the same cross section. It should 
be remarked that this phase uncertainty has nothing to do with the non-observable phase of wave functions in quantum 
mechanics; the asymptotic wave functions at large distances from the scattering centre may be written as $\Psi(x) 
\approx e^{i \cdot k \cdot x} + F(\theta) \frac{e^{i \cdot r \cdot }}{r}$, $ r \rightarrow \infty$, so the phase of 
scattering amplitude is the \emph{relative} phase of the incident and scattered wave. This phase has observable 
consequences in situations where multiple scattering occurs, and causes the continuum ambiguity. In the elastic region 
the unitarity relates real and imaginary parts of each partial wave, the consequence of the existence of equality 
relation is constraint which effectively removes the "continuum" ambiguity, and leaves potentially only a discreet 
one. The partial wave must lie \emph{on} the unitary circle. However, as soon as the inelastic threshold opens, 
unitarity provides \emph{only} an inequality: $ |1+2 \ i \ F_l|^2 \leq 1 \Longrightarrow {\rm Im F}_l = |F_l|^2+ I_l$, 
where $I_l = \frac{1}{4}(1 - e^{-4 {\rm Im} \delta_l})$. So each partial wave must lie \emph{upon or inside} its 
unitary circle, and not on it. A whole family of functions $\Phi$, of limited magnitude but of infinite variety of 
functional form satisfying the required conditions does exist, but in spite that they contain a continuum number of 
infinite points they are limited in extent. The ISLANDS OF AMBIGUITY are created. See Fig.1. 
\begin{figure}[h]
\centerline{\epsfxsize=2.5in\epsfbox{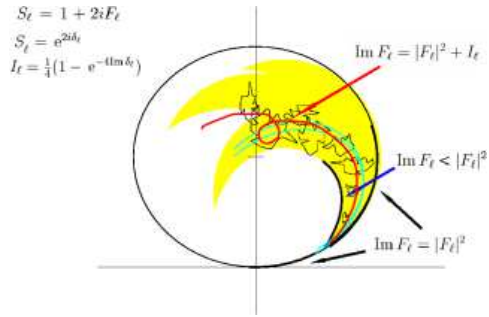}} 
\caption{ Creation of ISLANDS OF AMBIGUITY after the first threshold. }
\end{figure} \newline
Historically, the continuum ambiguity problem has been addressed from two aspects: a. as a mathematical problem 
(constraining the functional form $\Phi$) and b. as a physics problem (implementation of the partial wave T - matrix 
continuity; i.e. energy smoothing/search for uniqueness). In Bowcock/Burkhard\cite{Bow75} several smoothing schemes 
have been suggested and elaborated: i) the shortest path method; ii) explicit analytic parameterization in energy; 
iii) discrete ambiguities and energy dependence; iv) energy parameterization using dispersion relations; v) 
partial-wave dispersion relations and vi) fixed-momentum-transfer dispersion relations. For further study we refer the 
reader to this publication. 
\section{Continuum ambiguity and coupled channel formalism}
Let us formulate the way how we see the continuum ambiguity problem in the language of coupled channel partial wave 
analyses formalism (CC\_PWA). A commonly accepted postulate is that the T matrix is an analytic function of Mandelstam 
variables s and t. It is well known that each analytic function is fully defined with its poles and cuts. If an 
analytic function contains a continuum ambiguity it \emph{is not} uniquely defined in the whole complex energy plane, 
and the direct consequence is that we \emph{do not} possess a complete knowledge about its poles and cuts. \\
\emph{Conclusion:} \\
To eliminate continuum ambiguities in a coupled channel formalism approach it is essential to fully constrain T-matrix 
poles and cuts. \\
\emph{Basic idea:} \\
We want to demonstrate the role of inelastic channels in fully constraining the poles of the partial wave T-matrix, or 
alternatively said, we want to show their importance for eliminating continuum ambiguity which arises if only elastic 
channels are considered. \\
\emph{Implementation:} \\
Supplying scarce information for EACH channel is MUCH MORE CONSTRAINING 
 then supplying perfect information for ONE channel only. We shall use information from as many channels as possible 
in order to maximally constrain poles and cuts of T matrices in CC\_PWA. \\
In order to familiarize the reader with our concept of looking for poles in the complex energy plane we give: \\ 
 \emph{An attempt of a simple visualization} \\
 We are looking for a full set of poles of an analytic function in the complex energy plane while having at our 
disposal information originating from only restricted number of points on physical axes which we obtain by analyzing 
experimental data from elastic and inelastic processes. To illustrate our reasoning we present an analogue with a 
normal, everyday situation. Let us imagine that we are trying to get maximum information about a number of flour 
bouquets located on the table, not directly looking at them but having at disposal only their images in three mirrors 
located on three edges - Fig.2. Of course, the mirror closest to the particular bouquet will give the best information 
about it, it will take us much more effort (a magnifying glass for instance) to get the good information about 
bouquets further away. Sometimes, if two bouquets are located one behind another, we shall not be able to see the 
further one at all in some of the mirrors (the bottom right flower bouquets can be seen in the right mirror only). The 
bouquets in front will completely block our view. Of course, the only proper way is to look at all three mirrors at 
the same time. 
\begin{figure}[h] 
\centerline{\epsfxsize=2.5in\epsfbox{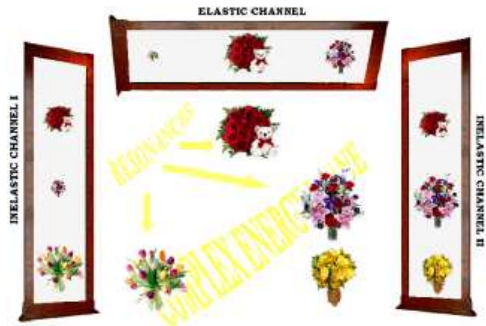}} 
\caption{The illustration of the search for resonances in coupled channel formalism with real world situation.}
\end{figure} \newline
The same goes for "looking for the resonances in the complex energy plane". \\
Let us identify flour bouquets with resonances, and mirrors with measurements in elastic and inelastic channels. The 
partial wave analyses which analyzes only elastic channel amplitudes will have to do a very good job to get good 
information about resonances lying far away from the real axis in the complex energy plane (the magnifying glass have 
to be very strong), and the information might be incomplete because one resonance might be masked. On the other hand, 
the formalisms which look at "all three mirrors" at the same time will have a better chance. 
We claim that the coupled channel formalism (looking at elastic and inelastic channels at the same time) is a method 
which reveals much more information about resonances then any procedure restricted to only one channel at a time, or 
differently said it is similar like looking at all three mirrors at the same time in the case of our flour bouquets at 
the table. 
\section{Coupled channel formalism}
For the collection of formulae we refer the reader either to original paper by Cutkosky et.al\cite{Cut79} or to one of 
the more recent CC\_PWAs; Zagreb\cite{Bat98} or Pittsburgh/ANL\cite{Vra00}, but to ease the understanding of the way 
how the conclusions are reached we give a flow diagram of a Carnagie-Melon-Berkeley (CMU-LBL) type formalism in Fig.3.
\begin{figure}[h] 
\centerline{\epsfxsize=2.5in\epsfbox{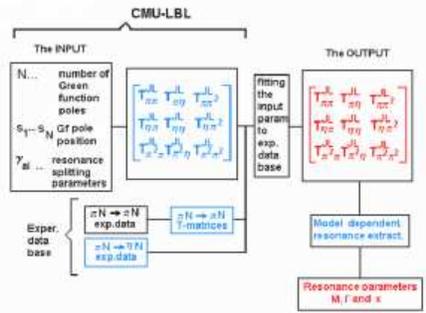}} 
\caption{The flow diagram for the Carnegie-Melon-Berkeley type formalism.}
\end{figure} 
\section{The N(1710)~P$_{11}$ resonance is a direct consequence of inelastic channels}
We shall use the afore introduced CC\_PWA formalism to illustrate how the presence of inelastic channel, $\pi N 
\rightarrow \eta N$ data in particular, is imposing the existence of the N(1710)~P$_{11}$ state. \\
\emph{Number of channels:} \\
We have simplified the problem in order to demonstrate the genesis of the new P$_{11}$ state by choosing the model 
with two channels only: $\pi N$ and the effective two body channel $\pi^2N$ which is representing all other two/three 
body processes in a form of a two body process with $\pi^2$ being a quasiparticle with a different mass chosen for 
each partial wave. \\
\emph{The data base:} \\
In principle we should fit experimental data. However, in that case all partial waves have to be simultaneously 
fitted, and the number of parameters becomes intolerably big. Instead, as the formalism separates individual partial 
waves, we choose to fit partial waves obtained directly from experiment so the fit can be performed with the reduced 
number of parameters. In other words, instead of using row experimental data, we choose to represent them as partial 
waves using \emph{any } form of partial wave analyses (single/multi-channel). The only criterion is that they indeed 
reproduce the experiment correctly. From that moment we treat the obtained results as the optimal amalgamation of 
different experimental data sets and regardless of their genesis use them as the experimental input.
\begin{figure}[h] 
\centerline{\epsfxsize=2.5in\epsfbox{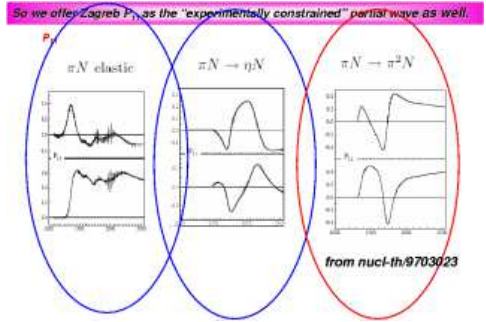}} 
\caption{The inelastic input used to illustrate the existence of the N(1710)~P$_{11}$ state.}
\end{figure}
 \newline
For P$_{11}$ pion-nucleon elastic partial waves we use the single energy VPI/GWU solution\cite{GWUWEB,Arn04}. \\
For representing the $\pi N \rightarrow \pi^2 N$ data in a form of the P$_{11}$ partial wave, we have followed the way 
it has been done in ref.\cite{Kis04}. The bottom line of the idea is that the number of experimental data in $\pi N 
\rightarrow \eta N$ channel is insufficient to perform a reasonable single channel partial wave analysis, so a kind of 
model should be introduced in addition. That is a coupled channel formalism. In that paper the Pittsburgh CC\_PWA 
results\cite{Vra00} are used as representing the \emph{experimentally} constrained S$_{11}$ T-matrix. Following that 
recipe, we have used the coupled channel curves from the analysis of Batini\' c et al\cite{Bat98a} where the world 
collection of $\pi N \rightarrow \eta N$ data is used to obtain partial wave T matrices, but instead of using the 
$T_{\pi N,\ \eta N}$ matrices which couple very poorly to the P$_{11}$ partial wave in the 1700~MeV range we have used 
the $T_{\pi N,\ \pi^2 N}$ part which is much stronger; see - Fig. 4. \\
 \emph{The procedure:} \\
 In order to show that fitting isolated channels results in different collection of T-matrix poles, we shall start by 
fitting channel by channel. We start with minimal number of intermediate particles, raise their number as long as the 
good fit is achieved, and then compare poles. If/when the collection of poles disagree (different collection of poles 
is needed to fit different channels), we fit all channels simultaneously until the quality of fit can not be improved 
by increasing the number of exchanged particles. \\ 
\emph{Elastic channel alone: } \\
Using two physical and two background poles we have fitted only elastic channel, and the obtained results are 
presented in Fig.5. Only one physical pole is sufficient to achieve the overall agreement of the model with the 
experimental input of ref.\cite{GWUWEB,Arn04}, and the pole is in the vicinity of 1400 MeV (Roper). Adding new poles 
is just visually improving the quality of the high energy end of the fit, and we can say that the existence of the 
second pole near 2100 MeV is only consistent with the data, and not required by them. Inelastic channel is reproduced 
extremely poorly. 
\begin{figure}[h] 
\centerline{\hspace*{-1.cm}\epsfxsize=4cm\epsfbox{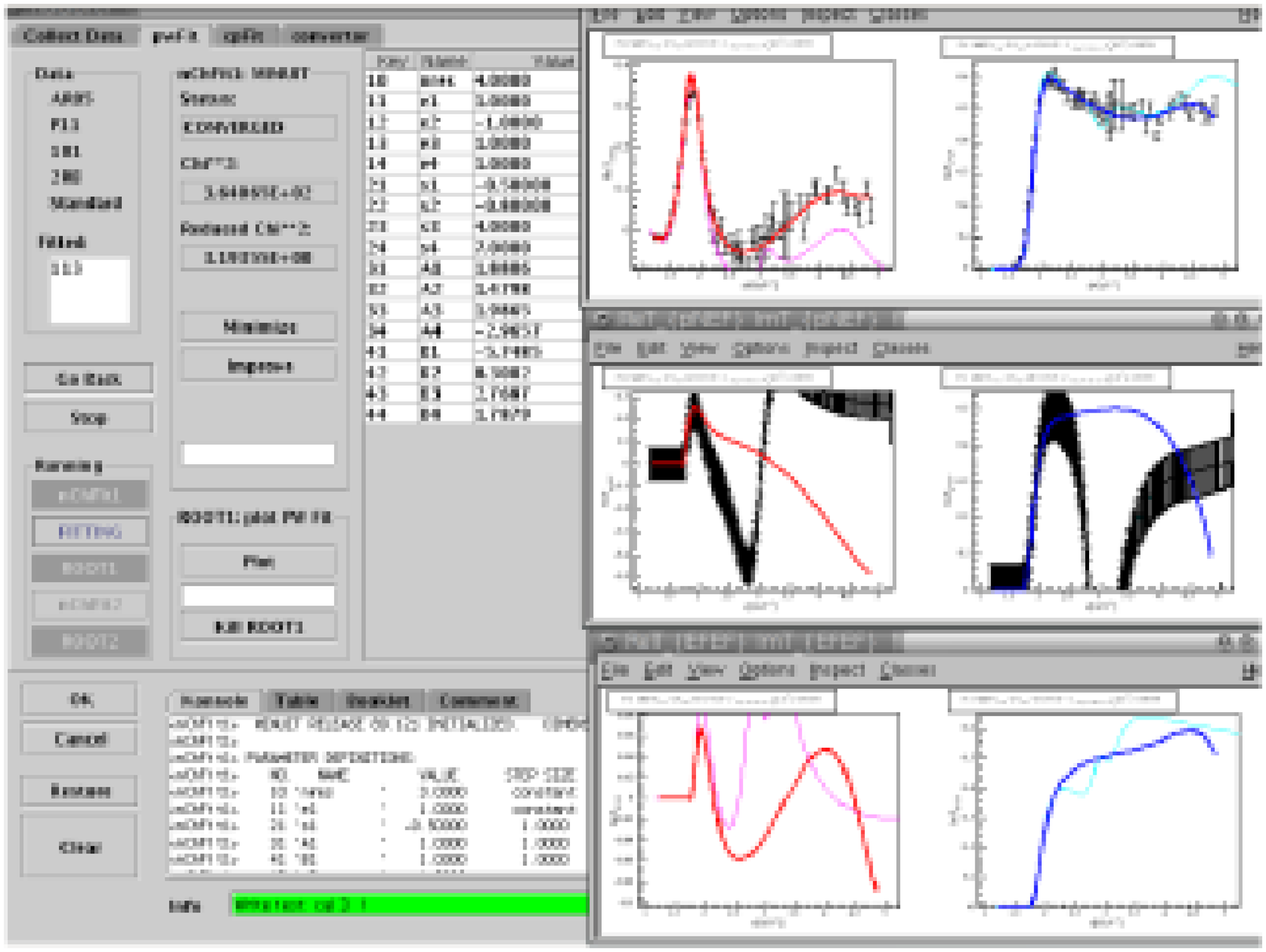} \hspace*{2.cm}
\epsfxsize=4cm\epsfbox{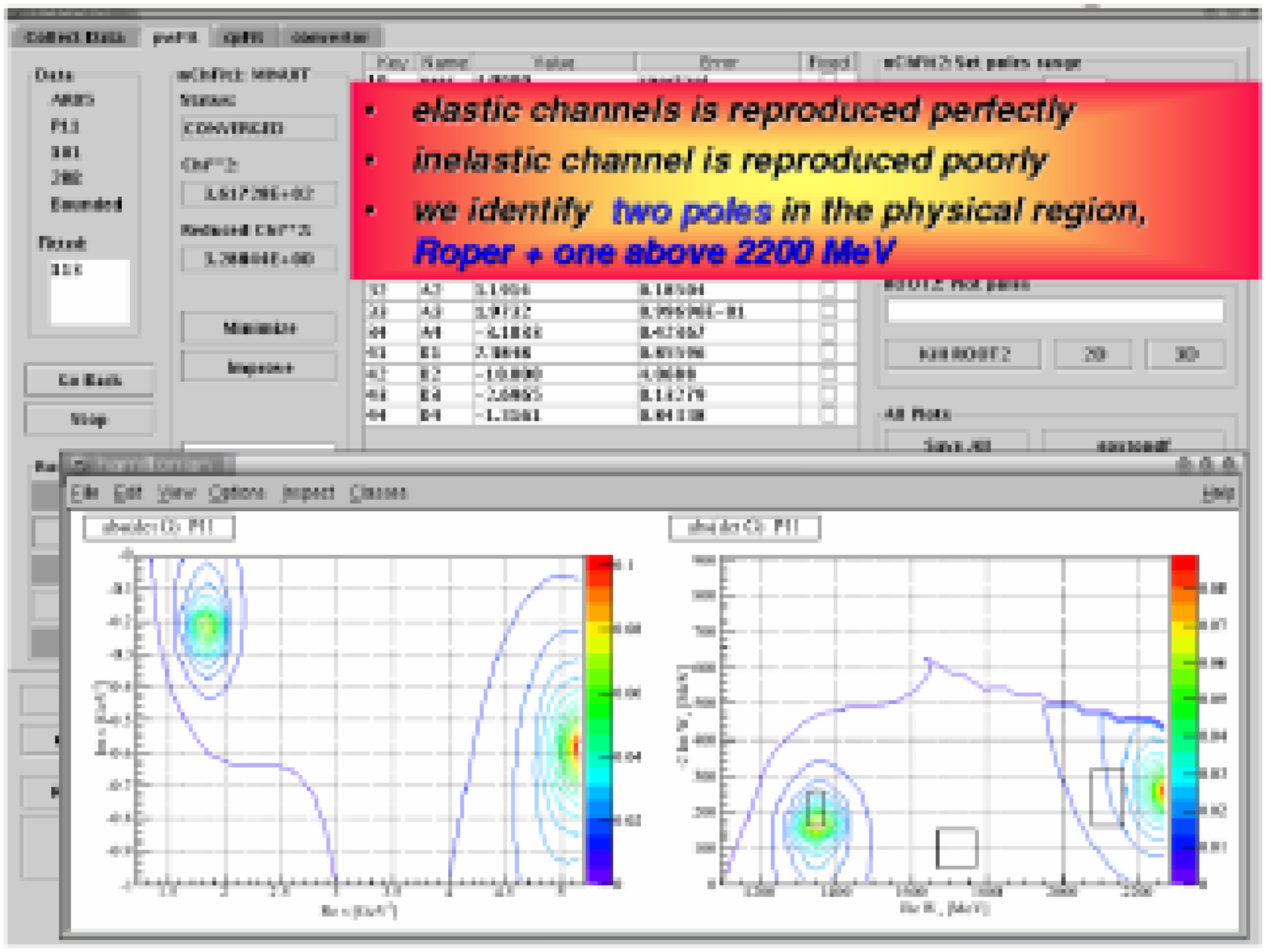} } 
\caption{The $\pi N \rightarrow \pi N$, $\pi N \rightarrow \pi^2 N$ and $\pi^2 N \rightarrow \pi^2 N$ P$_{11}$ 
T-matrices and pole positions for the two resonant/two background fit of elastic channel alone.}
\end{figure} \\ \\
\emph{Inelastic channel alone:} \\
Using two physical and two background poles we have fitted inelastic channel alone and the obtained results are 
presented in Fig.6. At least two physical poles are needed to achieve the overall agreement of the model with the 
experimental input of ref.\cite{Bat98a}, and poles are in the vicinity of 1400 MeV (Roper), and 1700 MeV. Elastic 
channel is reproduced extremely poorly. \vspace*{-0.3cm}
\begin{figure}[h] 
\centerline{\hspace*{-1.cm}\epsfxsize=4cm\epsfbox{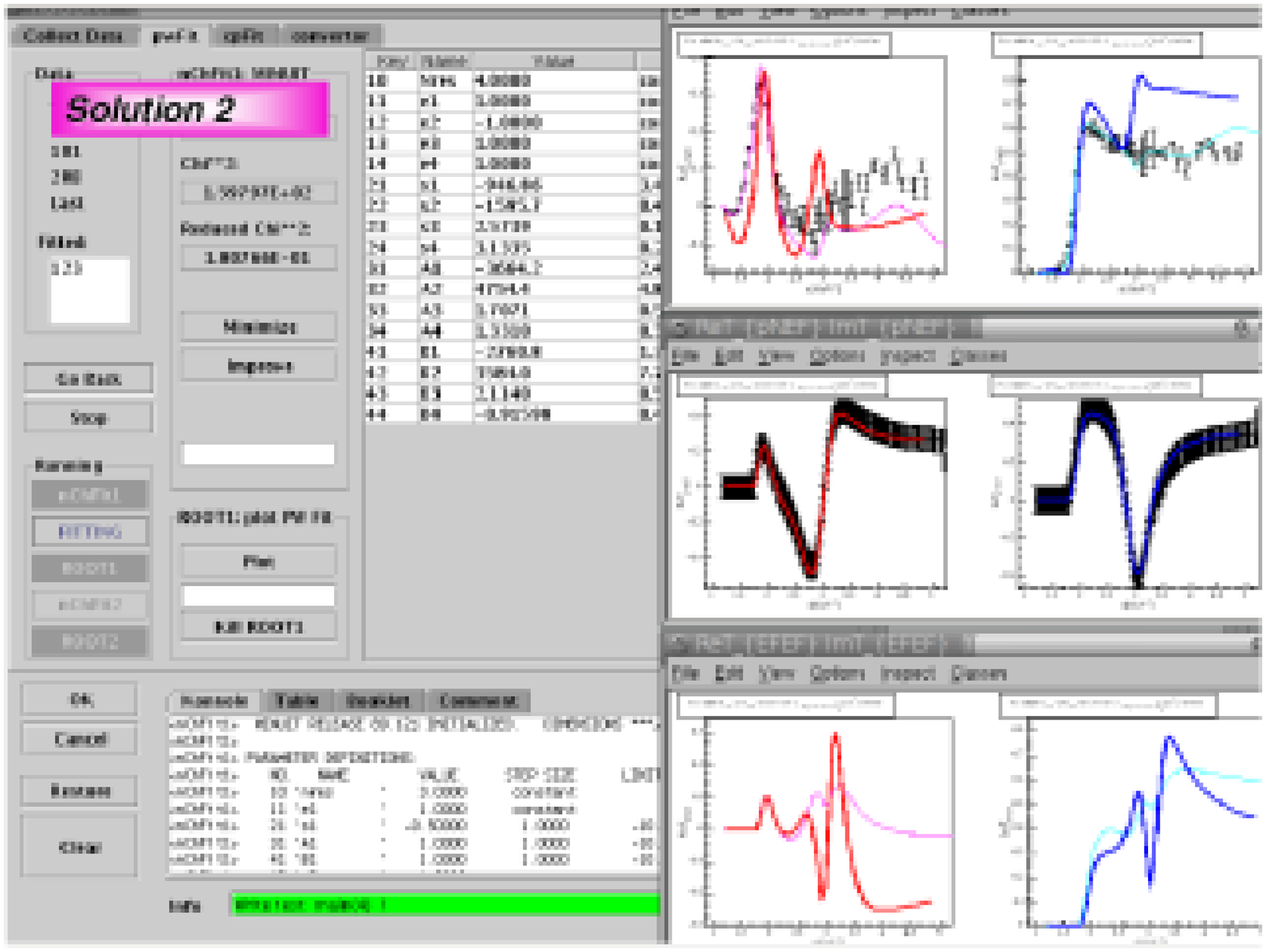} \hspace*{2.cm}
\epsfxsize=4cm\epsfbox{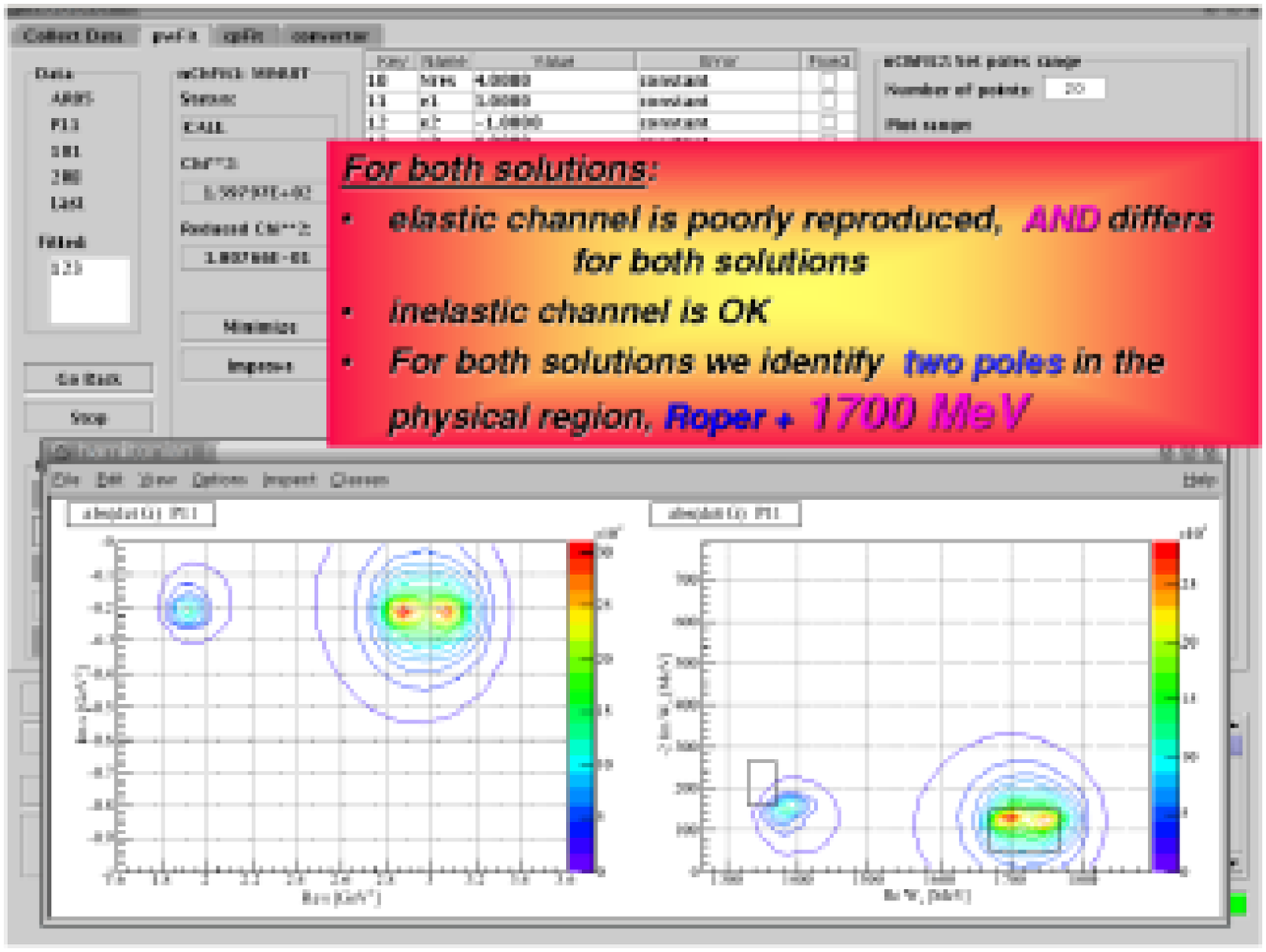} } 
\caption{The $\pi N \rightarrow \pi N$, $\pi N \rightarrow \pi^2 N$ and $\pi^2 N \rightarrow \pi^2 N$ P$_{11}$ 
T-matrices and pole positions for the two resonant/two background fit of inelastic channel only.}
\end{figure} 
\emph{Elastic + inelastic channel:} \\
Using two physical and two background poles we have simultaneously fitted elastic and inelastic channels and the 
obtained results are presented in Fig.7. At least two physical poles are needed to achieve the overall agreement of 
the model with the experimental input of refs.\cite{GWUWEB,Bat98a}, and poles are in the vicinity of 1400 MeV (Roper), 
and 1700 MeV. Both channels are reproduced. \vspace*{-0.3cm} 
\begin{figure}[h] 
\centerline{\hspace*{-1.cm}\epsfxsize=4cm\epsfbox{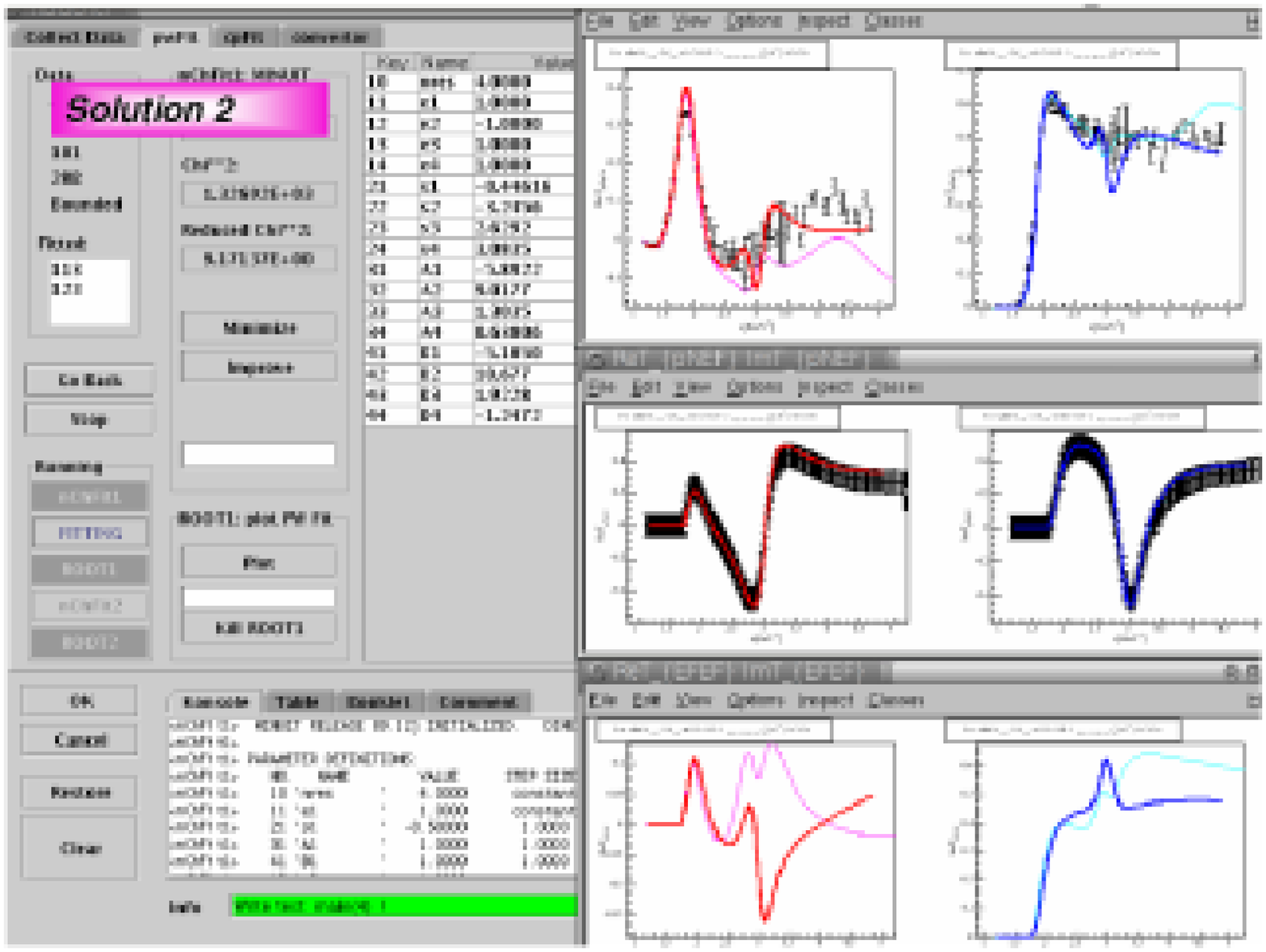} \hspace*{2.cm}
\epsfxsize=4cm\epsfbox{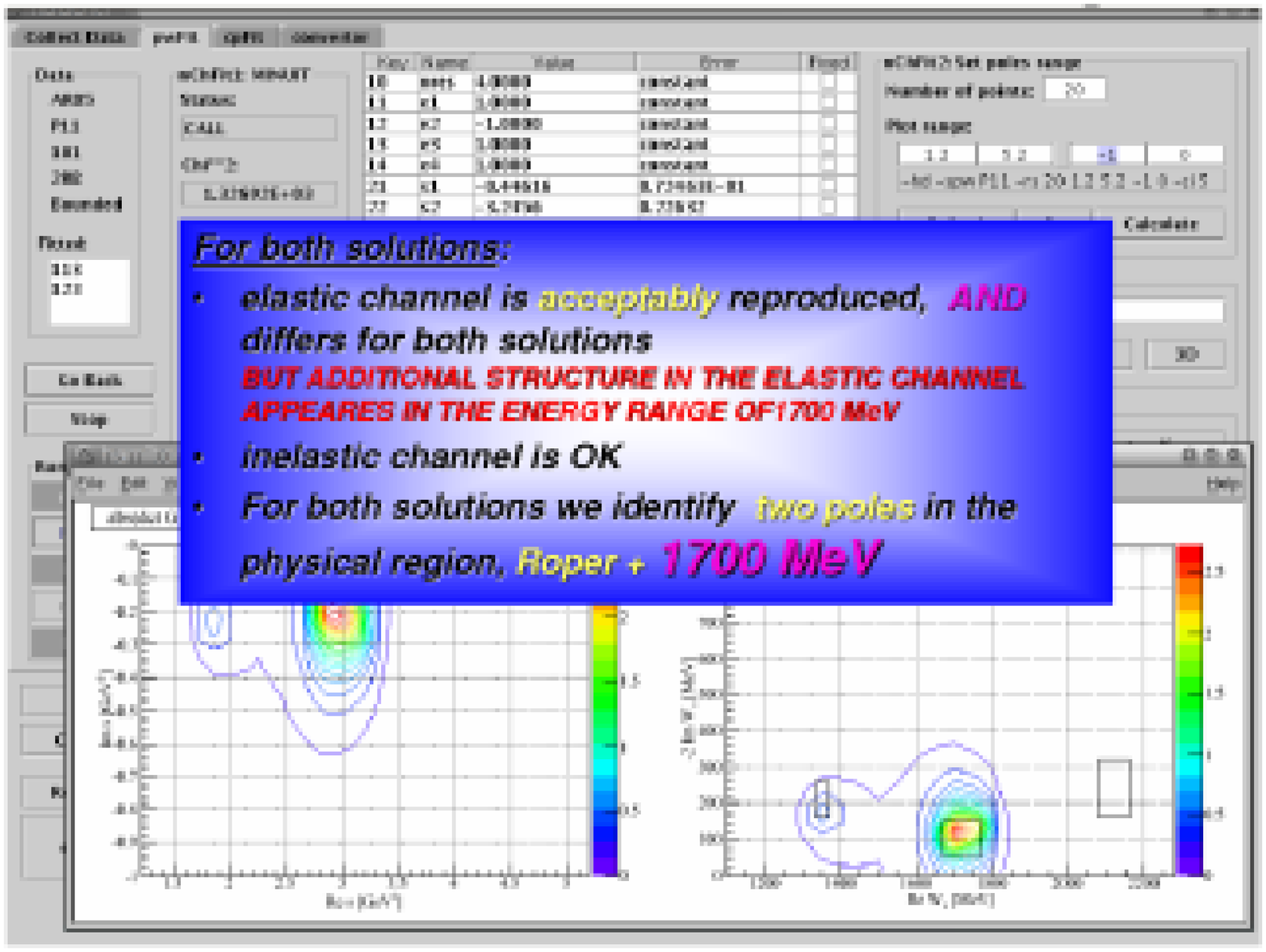} } 
\caption{The $\pi N \rightarrow \pi N$, $\pi N \rightarrow \pi^2 N$ and $\pi^2 N \rightarrow \pi^2 N$ P$_{11}$ 
T-matrices and pole positions for the two resonant/two background fit of both, elastic and inelastic channels.}
\end{figure} 
The simultaneous fit of elastic and inelastic channels requires that the energy behavior of $\pi N$ elastic T matrix 
is not as smooth as in FA02 solution of ref.\cite{Arn04}, but has an additional structure, very similar to solutions 
given in all old PWAs\cite{Cut79,KH84,Man92} and new ones\cite{Bat98,Vra00,Pen02}. 
However, single energy solution offered by the VPI/GWU PWA is consistent with the existence N(1710) P$_{11}$, because 
the structure not existing in FA02, and required by everyone else, can be understood as being hidden underneath 
disproportinably strong error bars which are reported in that solution in the vicinity of 1700 MeV - see Fig.8. \\ 
 \begin{figure}[h]
\centerline{\epsfxsize=2.5in\epsfbox{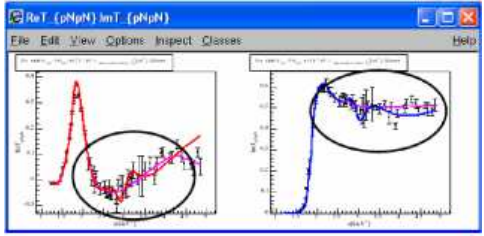}} 
\caption{ The $\pi N \rightarrow \pi N$ T matrix of FA02 (magenta line) and our solution (red and blue lines) for the 
two resonant/two background fit compared with SES of ref.\protect\cite{GWUWEB}}
\end{figure}   \vspace*{-0.3cm} 
\section{Conclusions}
T matrix poles, invisible when only elastic channel is analyzed, spontaneously appear in the coupled channel formalism 
when inelastic channels are added. \\
It is demonstrated that the N(1710) P$_{11}$ state exists, that the pole is hidden in the continuum ambiguity of 
VPI/GWU FA02, and that it spontaneously appears when inelastic channels are introduced in addition to the elastic 
ones. 
\section{Future prospects for utilizing inelastic channels data}
Instead of using raw data we propose to represent them in a form of partial wave T-matrices (single channel PWA, some 
form of energy smoothing can be as well introduced), and regardless of their genesis, use them in a CC\_PWA. The 
scheme if shown in Fig.9. 
\begin{figure}[h] 
\centerline{\epsfxsize=2.8in\epsfbox{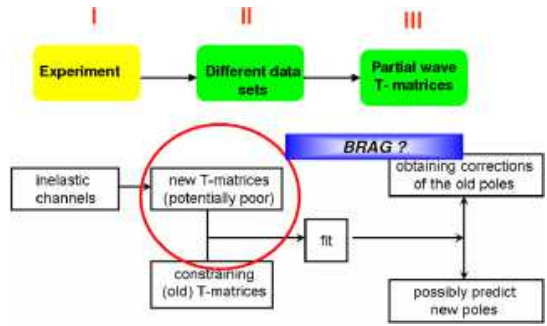}} 
\caption{The proposal for utilizing new inelastic data using the coupled channel formalism.}
\end{figure} \\
\emph{A call for help:} \\
Anyone who has some kind of partial wave T-matrices, regardless of the way how they were created please sent it to us, 
so that we could, within the framework of our formalism, establish which poles are responsible for their shape.

\end{document}